# Equivalent Time Active Cavitation Imaging

Simon Blais[1], Jonathan Porée[1], Gerardo Ramos-Palacios[2], Samuel Desmarais[1], Vincent Perrot[1], Abbas Sadikot[2], Jean Provost[1,3]

[1] Engineering Physics Department, Polytechnique Montréal, Canada
[2] Montreal Neurological Institute, McGill University, Montréal, Canada
[3] Montreal Heart Institute, Montréal, Canada

Information

Simon Blais: simon.blais@polymtl.ca

Jonathan Porée: jonathan.poree@gmail.com

Vincent Perrot: vincent.perrot@polymtl.ca

Gerardo Ramos-Palacios: gerardora11@gmail.com

Samuel Desmarais: samuel.desmarais@polymtl.ca

Abbas Sadikot: abbas.sadikot@mcgill.ca

Jean Provost: jean.provost@polymtl.ca

# Abstract

Rationale: Despite the development of a large number of neurologically active drugs, brain diseases are difficult to treat due to the inability of many drugs to penetrate the blood-brain barrier. High-intensity focused ultrasound blood-brain barrier opening in a site-specific manner could significantly expand the spectrum of available drug treatments. However, without monitoring, brain damage and off-target effects can occur during these treatments. While some methods can monitor inertial cavitation, temperature increase, or passively monitor cavitation events, to the best of our knowledge none of them can actively and spatiotemporally map the high intensity focused ultrasound pressure field during treatment.

Methods: Here we detail the development of a novel ultrasound imaging modality called Equivalent Time Active Cavitation Imaging capable of characterizing the high-intensity focused ultrasound pressure field through stable cavitation events across the field of view with an ultrafast active imaging setup. This work introduces 1) a novel plane wave sequence whose transmit delays increase linearly with transmit events enabling the sampling of high-frequency cavitation events, and 2) an algorithm allowing the filtration of the microbubble signal for pressure field mapping. The pressure measurements with our modality were first carried out in vitro for hydrophone comparison and then in vivo during blood-brain barrier opening treatment in mice.

Results: This study demonstrates the ability of our modality to spatiotemporally characterize a modulation pressure field with an active imaging setup. The resulting pressure field mapping reveals a good correlation with hydrophone measurements. Further proof is provided experimentally in vivo with promising results.

Conclusion: This proof of concept establishes the first steps towards a novel ultrasound modality for monitoring focused ultrasound blood-brain barrier opening, allowing new possibilities for a safe and precise monitoring method.

Abbreviations

BBB: Blood-brain barrier; ETACI: Equivalent Time Active Cavitation Imaging; FUS: Focused ultrasound; HIFU: High intensity focused ultrasound; PAM: Passive acoustic monitoring; PNP: Peak negative pressure; MRgFUS: Magnetic resonance imaging guided focused ultrasound

## Introduction

With the aging population, brain diseases are on the rise and are now a serious public health concern that needs to be addressed [1]. Brain diseases, such as neurological disorders or mass lesions, remain undertreated since most newly developed pharmaceutical agents that are active in vitro cannot access the brain due to the presence of the blood-brain barrier [2]. The BBB is a more selective filter [3] than the vasculature found in the rest of the body mainly due to the tight junctions between endothelial cells constituting the cerebral capillaries [4], [5]. Thereby, the BBB hinders more than 98% of neurotherapeutic drugs from acting on neurons and glia [6], [7]. Several approaches can currently overcome this barrier to reach the parenchyma, such as trans-arterial infusion of hyperosmolar agents [8], [9] and direct drug injection into brain tissue [10]. However, these methods to bypass the BBB are not optimal since they are either invasive or non-spatiotemporally specific [11] and can lead to adverse effects [12], [13]. More recently, a transient, localized, and non-invasive way of opening the BBB was made possible by combining microbubbles and High Intensity Focused Ultrasound (HIFU) [13]–[15]. This procedure overcomes the drawbacks of other techniques by utilizing the mechanical effect of microbubbles under ultrasound oscillations to promote anatomically precise opening in disease-relevant regions [16], [17]. The non-destructive nature of this method requires very low acoustic power and thus relies heavily on the microbubble's mechanical effect to achieve the transient opening [18].

Microbubbles are clinically approved contrast agents used routinely in ultrasound imaging to improve the detection of the vasculature [19]. The ease with which microbubbles are detected using ultrasound imaging comes from multiple phenomena including the impedance mismatch between the microbubbles and the surrounding environment, which generate strong backscattering of the incident imaging wave. Under the right ultrasound modulation condition, mechanical effects can be evoked to unsettle the integrity of the BBB by loosening tight junction complexes temporarily [15] while avoiding tissue damage. With the creation of low- and high-pressure zones in the tissue caused by the propagation of the HIFU pressure field in the medium, *in situ* microbubbles will undergo volumetric oscillations. Depending on the amplitude of the local pressure variation, the radial oscillations can take the form of stable or inertial cavitation. At low peak negative pressure (PNP), the volumetric oscillations of the microbubbles are stable around their equilibrium radius at rest, leading to apparent damage-free, reversible BBB disruption [17], [20], [21]. On the other hand, at higher PNP, stable cavitation gives way to inertial cavitation, which is the result of unstable radial oscillations. Above a certain pressure threshold, the unstable oscillations result in a violent collapse of the microbubbles, damaging the vasculature and brain tissue [22], [23]. One way to distinguish the two types of cavitation is by their backscattered acoustic emissions under ultrasound modulation [24]. Stable volumetric oscillations of the microbubbles lead to strong backscattered linear and nonlinear acoustic emissions, such as harmonics, sub-harmonics, and ultra-harmonics of the transmit HIFU central frequency [25], [26]. As the pressure amplitude increases, the non-linearity of the signal becomes more important until the previously mentioned pressure threshold is reached, leading to unstable growth and collapse. Upon collapsing, energy is released as heat and broadband acoustic emissions [27], [28]. The pressure threshold at which inertial cavitation occurs depends on multiple factors, such as bubble size, local *in situ* temperature, and the excitation frequency [29] and goes to show how precarious [30] the state of the cavitation can be without proper monitoring. Given that the main objective of treatment for major brain diseases is to open the BBB in a controlled, damage-free manner, inertial cavitation must be avoided.

HIFU-induced BBB opening is arduous and would ideally require real-time spatiotemporal monitoring to ensure that the treatment location and the pressure distribution are as planned. Furthermore, multiple bioeffects take place *in vivo* that can shift the HIFU focal zone from the planned target. Thermal lensing, tissue motion, and beam distortion can displace the focal field relative to the theoretical ultrasound focal field [31]–[34]. While the effect is negligible in small animal models such as mice, they become more pronounced in larger species [35], [36]. For these reasons, spatiotemporal HIFU pressure field mapping is mandatory for safe and effective BBB opening in clinical contexts. The current gold standards that are clinically in use for focused ultrasound (FUS) treatment monitoring include either MRI-based or ultrasound-based methods [37]. The first consists of Magnetic Resonance Imaging (MRI) thermometry measurements [26] that can map temperature variation of around 0.3°C [38] to localize the FUS focal point at a millimeter spatial resolution [39] with a temporal resolution in the order of a second [38]. However, there are still several issues with MRI-guided FUS (MRgFUS), as it can't monitor acoustic emissions, lacks portability, and requires MRI-compatible transducers. Moreover, BBB opening does not necessarily involve temperature increases and requires tight monitoring of acoustic emissions, thus making MRI monitoring ineffective for this purpose. One approach to localizing the focal zone of *in situ* during MRgFUS BBB opening is to first induce a minor increase in temperature to visualize its position with thermometry moments before starting the procedure [13] at the lower PNP amplitude for BBB opening. However, this approach lacks real-time feedback once the procedure is started and is susceptible to misalignment due to tissue motion or other beam distortion problems as the treatment progresses. Also, it is important to keep in mind that the majority of the patients evaluated for FUS treatments are elderly, and lengthy MRgFUS procedures that can extend over more than 3h limit tolerability [40], [41]. Post BBB opening assessment can also be done to reduce the time in the MRI, but this approach also lacks the real-time monitoring aspect. To address the issue of cavitation, a passive cavitation detector (PCD) is regularly added to the system to monitor the acoustic emission of the cavitating bubble during FUS procedures.

Some ultrasound-based methods are used to visualize inertial cavitation events that create hyperechogenic regions in the tissue during FUS procedures that use output pressure in the order of multiple MPa [42], which is well above the BBB opening pressure range and thus not suitable for this application. Recently, various ultrasound monitoring techniques have been developed, including passive acoustic monitoring (PAM) based methods. PAM techniques use the FUS transducer as the emitter while a receiver array passively records the backscattered signal. PAM techniques are extensively characterized and are an improvement compared to conventional PCD monitoring, which has a poor spatiotemporal resolution as it can only detect cavitation events from a fixed volume and can lead to misinterpretation since signals from acoustic events can interfere with each other [43]. Spatial resolution can be increased with the use of linear arrays and passive beamforming methods [37], [44] but given that no imaging pulse is emitted, the axial resolution is suboptimal.

Simulation is also a powerful tool to help visualize the FUS pressure field. However, simulation remains unreliable for *in vivo* treatment prediction as the FUS BBB opening is based on the interaction between the FUS pressure field, the tissues, and the microbubbles. Simulation methods therefore often lack information of important biological characteristics including brain tissue properties, blood flow, and spatial microbubble concentration [45]. Furthermore, accurate simulation requires pre-op imaging such as CT to extract the geometry and the density of the patient's skull [41] to reproduce the aberrations, thus adding extra steps to the procedure.

Despite the important attempts in addressing the drawbacks such as spatial and temporal resolution, pressure range, and real-time monitoring, none to our knowledge can spatiotemporally monitor the FUS pressure field by actively mapping stable cavitation events. Given the importance of achieving reliable and controlled BBB opening and the limitations of current approaches, there is a pressing need for a low-cost, high-resolution spatiotemporal pressure field mapping solution.

Over the last few years, ultrasound imaging has made major improvements with the advent of ultrafast programmable ultrasound scanners. High-speed ultrasound acquisition and accumulation of large data sets have allowed the development of new imaging modalities, making it possible to image inaccessible structures with other non-invasive imaging modalities [46]. Here we combine these new approaches, to present a proof of concept of a novel ultrasound imaging modality, Equivalent Time Active Cavitation

Imaging (ETACI), that can monitor the FUS pressure field by utilizing an active imaging setup and the radial oscillation of microbubbles under sonication. This method allows for adjustability in the parameters, such as frequency and pressure for both the imaging and the therapy emission and has a good resolution in the axial dimension given the active imaging nature of this method. The goal of this study is not to quantitatively measure acoustic pressure field but to have visual guidance of the localization of the pressure field in relation to the biological structures in the field of view.

## Methods

To map the pressure field of the HIFU transducer, we developed a modality where we can isolate backscattered signals of stable cavitating microbubbles from the rest of the tissue using an active imaging setup. As the microbubble is submitted to pressure waves propagating through the medium, it will undergo cavitation. High- and low-pressure zones will induce a volumetric oscillation of the microbubble, which will affect its backscattered signal when hit by an imaging pulse [47]. When the microbubble is in a rarefaction zone, the diameter of the bubble will be larger than at its equilibrium states, and conversely for when the bubble is in a compression zone (Figure 1). When the microbubble is larger than its equilibrium state, the backscattered signal will be higher than when the volume is below the equilibrium state [48], [49]. The reflected signal will thus depend on the volumetric oscillation state of the microbubbles which in turn is related to the local pressure field intensity. Due to this phenomenon, the microbubbles will therefore serve as pressure sensors [50] throughout the medium and will indicate whether, at a given location, the medium is subjected to the pressure field of treatment. Knowing that the backscattered signal of the tissue over time should stay constant [48], we can thus differentiate the microbubble from the tissue by the variability of the signal in the slow time dimension. To reliably measure the treatment pressure output, it is critical to image simultaneously with FUS treatment as oscillation damping will affect any delayed measurements [51]. Plane-wave emission was used for this modality because, unlike focused imaging, the additional pressure modulation induced by the imaging probe avoids the risk of microbubble disruption. It was also used for coherent temporal mapping across the whole field of view as biological effects occur. Ultrafast plane wave imaging also increases the number of images that can be acquired in a defined time frame, making this method suitable for real-time monitoring. In this study, there is an underlying assumption that the microbubbles mostly oscillate symmetrically. However, this simplification might not be true in an in vivo environment where microbubbles oscillation can be obstructed when positioned near a blood vessel wall.

### Equivalent Time Active Cavitation Imaging sequence

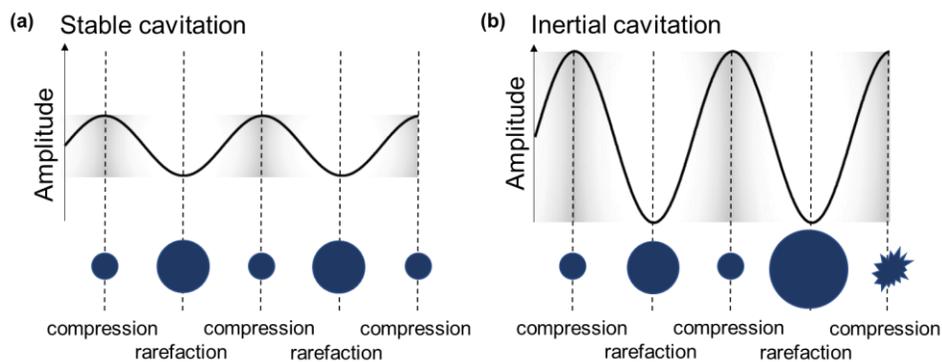

*Figure.1 : Microbubble's volumetric oscillation under modulation wave propagation through the medium. (a) Stable cavitation resulting in cyclic volumetric expansion and compression. (b) Inertial cavitation resulting in microbubble collapse.*

As we are limited by the round-trip time needed for the signal to be emitted and received from the medium, it is not possible to actively sample the microbubble oscillation at frequencies that are high enough to respect the Nyquist criterion. To bypass these limitations, ETACI is based on the use of a time delay in the emission of the imaging pulse equivalent to a fraction of the period of the modulation wave (Figure 2). By doing so, we interrogate the medium at different equally spaced time frames relative to the modulation wave period (Figure 2(a)). We assumed that the oscillation of the stable cavitating microbubble should be periodic in time and therefore, should be equivalent when spaced by an integer of modulation cycles. The transmit delay is calculated by the number of wavelengths of the imaging probe central period (1/Imaging central frequency) and is determined beforehand. A custom DAS beamformer was adapted to correct for the added delay of the imaging pulse to register the images spatially onto the same grid.

For this feasibility study, the treatment transducer was a single element high-intensity focused ultrasound transducer (HIFU H-102, Sonic Concepts, WA) driven at 1 MHz. The number of sampled microbubble oscillation states over a full HIFU period $P_{HIFU}$ is defined by the integer $k$. The number of full HIFU periods $n$ between every acquisition dictates the imaging frequency (1/PRP).

$$PRP = n * P_{HIFU} \qquad \Delta_t = \frac{P_{HIFU}}{k} \qquad (1)$$

When the time between two acquisitions is an integer number of HIFU periods, the microbubbles are sampled at the same volumetric state, which translates to a constant backscattered signal along the slow time dimension (Figure 2(b-c)). The microbubble signal is thus clustered with the tissue signal around the null frequency. As we introduced the time shift in between acquisitions, we sample the microbubbles in a series of modulated states, which creates a periodic oscillation in the back-scattered signal of the imaging pulse along the slow time (Figure 2(d)). This signal intensity oscillation along the slow time domain translates to a shift in the frequency domain (Figure 2(e)), making the microbubble signal stand out of the clutter signal. Thus, by varying the induced time shift in the imaging sequence, it is possible to modify the frequency spectrum of the back-scattered emission in slow time to isolate the signal of interest in a controlled manner. We assume that unless inertial cavitation or the absence of microbubbles occurs over time, the volumetric oscillation signal should be periodic along the slow time. This method is thus only valid under low-pressure conditions, which is a limitation of the study, and it is consequently assumed that the microbubbles are driven in a linear regime. It is thus possible to subtract the PRP time in between acquisitions as it is an integer number of HIFU periods and thus obtain an equivalent sampling period of Δt instead of PRP+Δt (Figure 2(f-g)). By doing so, we were able to artificially sample high-frequency events in the slow-time dimension higher than the possible limit of ultrafast ultrasound imaging. This type of processing has some similarities with SURF imaging [48], uRMI [52], and Equivalent-Time Sampling, which is a sampling method implemented in a digital oscilloscope. The results presented in this paper were acquired with k= 5 and an imaging period of 500 μs.

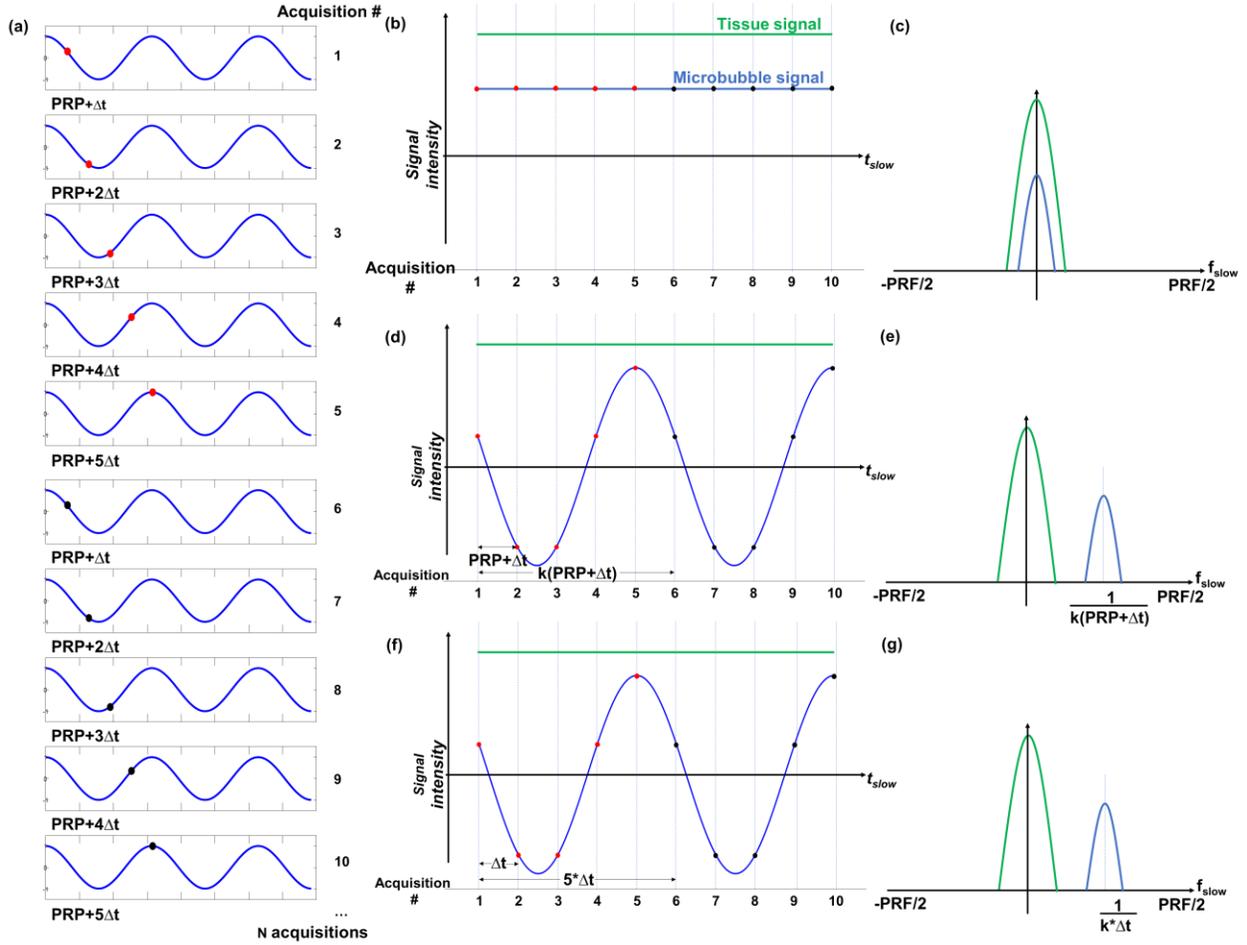

*Figure 2 : Overview of the ETACI sampling and its effect on the received signal. (a) Sequence acquisition shifts in relation to the modulation wave. (b) Schematic of the back-scattered signal of the microbubbles and the tissue when the microbubble is sampled periodically in the same volumetric state. (c) Frequency spectrum in the slow-time of the backscattered signal. (d) Schematic of the back-scattered signal of the microbubbles and the tissue when the sequence samples the microbubbles at various states along a volumetric oscillation cycle. (e) Frequency spectrum in the slow-time where the microbubbles' signal is shifted from the clutter signal. (f) Schematic of the back-scattered signal of the microbubble and the tissue when the sequence samples the microbubble in various states along a volumetric oscillation cycle and the equivalent time relation. (g) Frequency spectrum in the slow-time where the microbubbles' signal is shifted from the clutter signal and the equivalent time relation.*

## Pressure field map extraction

A Fourier analysis is performed for all corresponding pixels of the reconstructed images along the slow-time. Depending on the frequency of interest, the corresponding Fourier coefficient can be selected, resulting in an intensity map. The magnitude of the Fourier coefficients in each pixel characterizes the amplitude of the sinusoidal signal at the selected frequency in the backscattered signal and this value can then be translated and interpreted in pressure intensity, thus characterizing the pressure field intensity across the field of view. In this study, we used up to 250 ensembles of 250 frames to generate compounded pressure intensity maps. The simplicity of this method facilitates the implementation for real-time monitoring and can be tailored to different modulation frequencies and imaging probes.

## Transducer Setup

For this study, a dual transducer setup was designed to allow *in vitro* testing as well as *in vivo* testing with a view to eventual clinical translation (Figure 3). A custom, 3D-printed setup was fabricated to mount a CL15-7 ultrasound probe (ATL, WA) concentrically to the central opening of the HIFU transducer (H-102, Sonic Concepts, WA). The central frequency of the chosen transducers was 8.9 MHz and 1.1 MHz for the imaging probe and H-102 respectively. The 3D printed setup also served as an attachment point between the 3D positioning system and the dual transducer montage. To facilitate the fitting and positioning of the irregularly shaped CL15-7, the probe was cast in silicone (Mold-Star 30, Smooth-on, PA) and then fit in the 3D printed holder. A rubber ring was compressed in between the holder and the casing of the H-102 to allow the water to rise out of the coupling cone and submerge the imaging probe. As the imaging probe measures 36.5 mm and the opening of the H-102 measures 20 mm, we were limited to an aperture composed of 88 out of the 128 elements to transmit and receive information from the imaging probe. When in a coaxial position on top of the modulation transducer, the imaging array was sitting at around 72 mm from the exit plane of the coupling cone. The bottom opening of the coupling cone was sealed off with a latex membrane (Durex, USA) and a rubber O-ring. The water was circulating in the coupling cone in between acquisitions using a water degassing system (WDS-105+, Sonic Concepts, WA). The ultrasound probe was connected to a 256-channel Vantage ultrasound system (Verasonics, WA). The HIFU input was connected through a matching impedance box (Sonic Concepts, WA) that was fed with a 1 MHz sinusoidal output from the function generator (32220A, Agilent, USA) that had been amplified by a power amplifier (BBS0D3FOQ, Empower RF System, CA) at 40 dB.

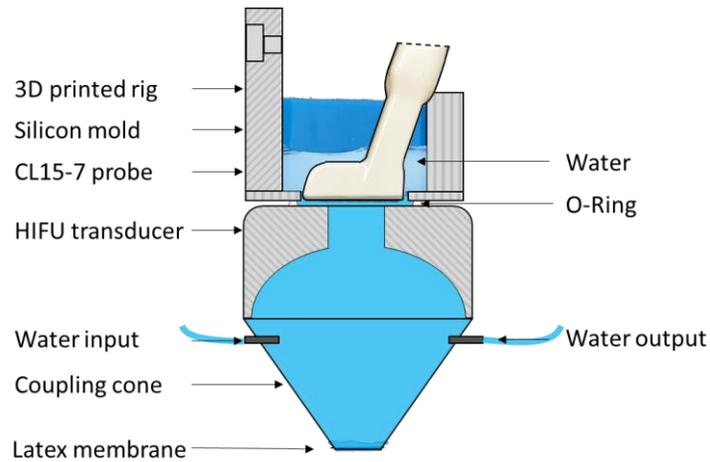

*Figure 3 : Co-axial transducer setup for blood-brain barrier opening treatment in mice.*

## *In vitro*

### Free field Characterization

The transducer montage was submerged halfway in a degassed [$O_2 \approx$ 5 mg/L, T $\approx$ 22 °C] 12 Liters water tank containing $2.4 \times 10^7$ Definity microbubbles (Lantheus Medical Imaging, USA) aiming at an ultrasound absorber laying at the bottom of the water tank. Continuous focused ultrasound exposure was applied for 120 seconds while imaging simultaneously with ETACI. The CL15-7 was operated in transmit/receive mode and for each acquisition, 50 repetitions of 5 consecutive delay-induced frames were acquired, for a total of 250 frames. The frames were accumulated into an ensemble, then transferred and saved to a solid-state drive for offline processing. This was repeated for a total of 250 ensembles and all the received data were post-processed offline on MATLAB (MathWorks, USA). The resolution of the beamformed images was λ/4 in both axial and lateral directions. The objective of this experiment was to have a clear unobstructed view of the pressure field to compare it to direct measurement using a hydrophone. Also, to characterize the effect of microbubble concentration and pressure on the signal amplitude, we first compared the signal spectrum of microbubble oscillation under the same pressure modulation, but with different microbubble concentrations. Next, varying pressure amplitudes were applied with a constant microbubble concentration. Since microbubbles decay in few minutes [53], we ensured a constant microbubble concentration between each test by flushing the water tub, refilling and degassing, before reinjecting microbubbles in the tub for sonication.

### Hydrophone mapping

Hydrophone mapping has been performed using the same conditions as the free field imaging. spatial pressure mapping was achieved with three motorized linear stages (X-LSM200A-E03, Zaber, CA) and a hydrophone (Y-104, Sonic Concepts, WA) in a 160 L custom water tank. The HIFU was targeted at the hydrophone for 100 cycles in burst mode and the output signal from the hydrophone was acquired with a data acquisition card (Picoscope 5242D, Pico Technology, UK) at a sampling rate of 8.93 MHz. After each measurement, the HIFU was moved in space to characterize the full pressure field of the center plane of the focal zone. The 60 middle cycles acquired by the acquisition card were then used to quantify the minimal voltage at each pixel. That voltage was then converted to the negative pressure using a calibration factor of 12 V/kPa. The maximal negative value at the focal zone was used to define the PNP. All the different communication protocols to interact with the different subsystems were implemented in Python.

### Static Phantom Characterization

To characterize the capability of tissue suppression of this method, *in vitro* testing with a tissue-mimicking phantom (Blue Phantom, CAE Healthcare) and contrast agent was performed (Figure 4). The transducer setup was submerged in a degassed 4 L water tub [$O_2 \approx 5$ mg/L, T ≈ 22 °C] containing 1.2 x $10^8$ free-floating Definity microbubbles and a phantom was placed less than halfway in the field of view. Continuous focused ultrasound exposure of 0.025 MPa PNP was applied while imaging with our custom plane sequence for 250 ensembles.

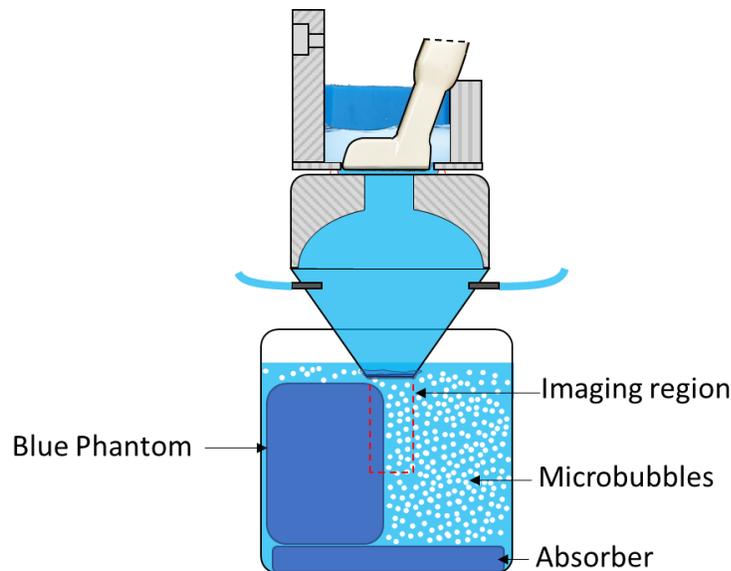

*Figure 4: Co-axial transducer setup for in vitro characterization of the method.*

### Flow Phantom Characterization

A flow phantom setup was assembled to characterize the capability of mapping the pressure field in a vessel and the effect of fast-moving microbubbles. *In vitro* testing with a tissue-mimicking flow phantom (HM 2.0, Humimic Medical, USA) and $2.4 \times 10^7$ Definity microbubbles diluted in 100 mL was performed with the gravity-generated flow. A ~5 mm vessel was created in the phantom just under the surface so it would go through the imaging region and across the focal region. The phantom was submerged in water and placed above an absorber in order to reduce reverberation. The transducer was in direct contact with the flow phantom and a layer of ultrasound coupling gel was laid in-between the two interfaces for better ultrasound coupling. The 1 MHz FUS output was set to 0.025 MPa PNP and the acquisition lasted for 250 ensembles.

### *In vivo*

One female wild-type mouse (C57BL/6J, Jackson Laboratories) underwent FUS treatment for BBB opening. Total treatment duration was 120 seconds and was aimed at the right or left striatum consecutively, using either continuous or burst sonication respectively. The animal was prepared beforehand, and the scalp fur was shaved and denuded with a depilatory cream. The transducer pressure output was set to 0.40 MPa PNP for either side and a duty cycle of 20% over a 50 ms period was used for the burst treatment. The targeting of the striatum was laser-guided on the scalp with a 3D positioning system. Around 10 seconds before sonication, a cocktail containing 25 µL of microbubbles ($1.2 \times 10^8$ Definity microbubbles in 0.98 mL of saline) and Evan's Blue (1% Evan's Blue in PBS at 7.4 pH at 2 µl/g of mouse body weight) was injected into the tail vein. For ultrasound imaging, 150 ensembles of 250 images each were acquired simultaneously during the treatment, with a 2000-Hz pulse repetition frequency. A 5-minute pause was applied between the two sonications in order to let the microbubbles fully decay and a second bolus injection was then administered through the tail vein. A small number of ensembles containing mostly white noise, probably caused by electromagnetic interference from the other devices in the *in vivo* experimental set-up were removed for the creation of the compounded ETACI maps.

Freeze-mount fixed brain sections (40 microns thick) were tiled under a fluorescent microscope. The spatial registration overlay between a coronal brain tissue section and the ETACI pressure field mapping was generated using B-mode images (Figure 9(c-f)) and stereotactic coordinates.

## Results

### *In vitro*

#### Free field mapping and hydrophone correlation

ETACI is intended to show the intensity of the pressure field through the intensity of pixels visualized in the resulting filtered image. However, as the microbubble concentration increases, the pixel intensity also increases under the same modulation conditions. As shown in Figure 5(a), the microbubble concentration influences the amplitude of the signal around the frequency of interest. This condition makes the following tests harder to accomplish because of the inherent effect of the microbubble concentration on the results, which need to be precisely standardized in between testing conditions to see the pressure/intensity relation without a microbubble concentration bias. Figure 5(b) shows that ETACI is sensitive to changes in the intensity of the modulation wave PNP. As intended, we observe a relationship between the amplitude of the microbubble oscillation through the value of the Fourier coefficient and the HIFU pressure output at the frequency of interest. The characterization in the free field allows us to find the frequency of interest of cavitating microbubble and to determine the response of the modality when no modulation wave is sent through the medium. As Figure 5(b) shows, the spike around the 1-MHz band is constant regardless of the modulation intensity, indicating that the microbubble oscillation frequency does not correspond to the expected frequency of 1 MHz [51], which is HIFU driving frequency. The volumetric oscillation frequency of the microbubbles corresponds to a signal around the 1.12 MHz band, and the intensity of the corresponding signal depends on the modulation transducer PNP output. For all the following results, the frequency band of interest is chosen to be between 1.11 MHz to 1.12 MHz.

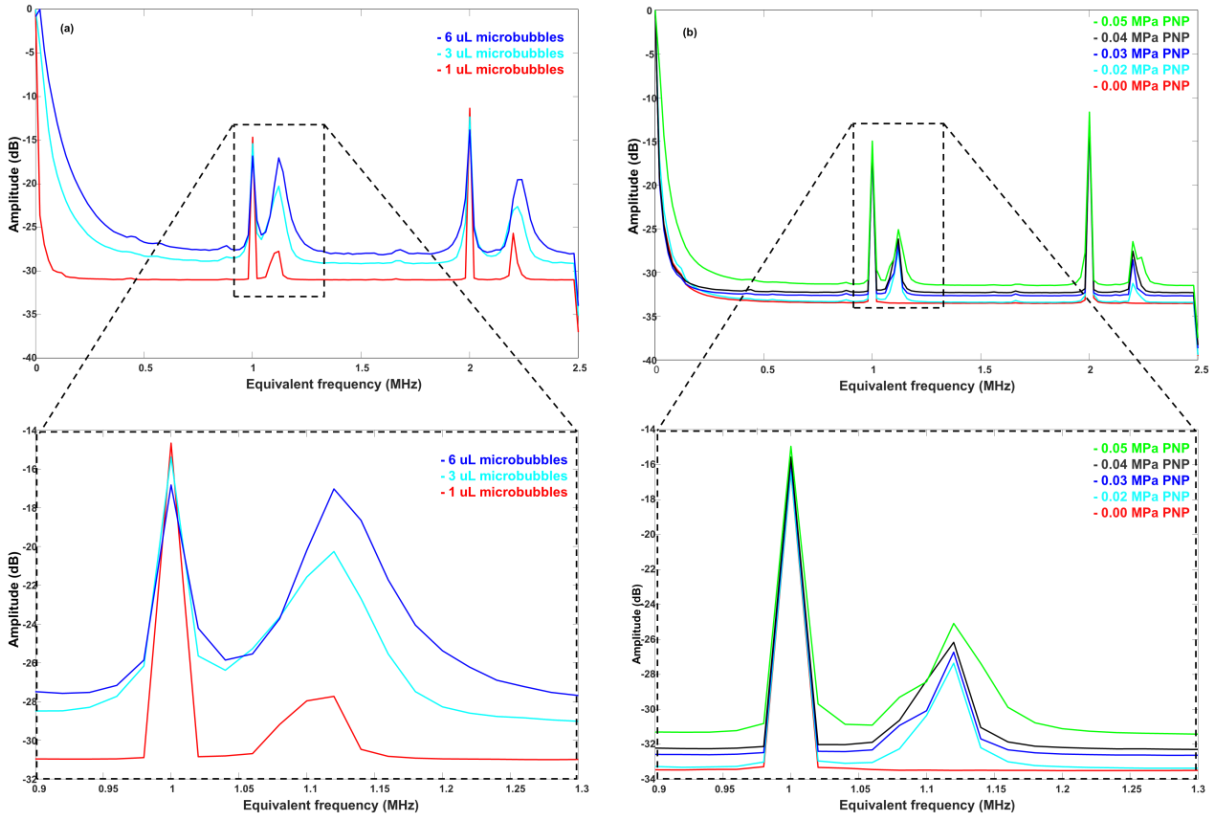

*Figure 5: Mean slow-time frequency spectrum at the equivalent sampling frequency for every pixel of the field of view (FOV). (a) Spectrum profiles at different microbubble concentrations under the same high intensity focused ultrasound modulation (0.1 MPa). Zoomed in spectrum between 0.9 MHz and 1.3 MHz. (b) Spectrum profiles at variable high intensity focused ultrasound peak negative pressure amplitude with a constant microbubble concentration. Zoomed in spectrum between 0.9 MHz and 1.3 MHz.*

To evaluate the validity of the pressure mapping using our method, we compared the resulting compounded map to the pressure mapping obtained using hydrophone measurements performed in the midplane of the focal field. Figure 6(a) shows the overall FOV starting at 77 mm from the imaging probe with the hourglass-like shaped pressure field. Figure 6(d) shows the close resemblance in-between the mean profiles measurements of the two mapping techniques (Figure 6(b-c)) along the HIFU axial direction. Both these profiles show similar dips around the central peak as expected for a FUS focal pressure profile. The dependency obtained between the corresponding pixel intensity value of ETACI and the measured pixel negative pressure shown in Figure 6(e) gives a linear correlation with an R-square of 0.7906. In order to have the same number of data points between the two datasets, the ETACI image was down sampled to the size of the hydrophone map.

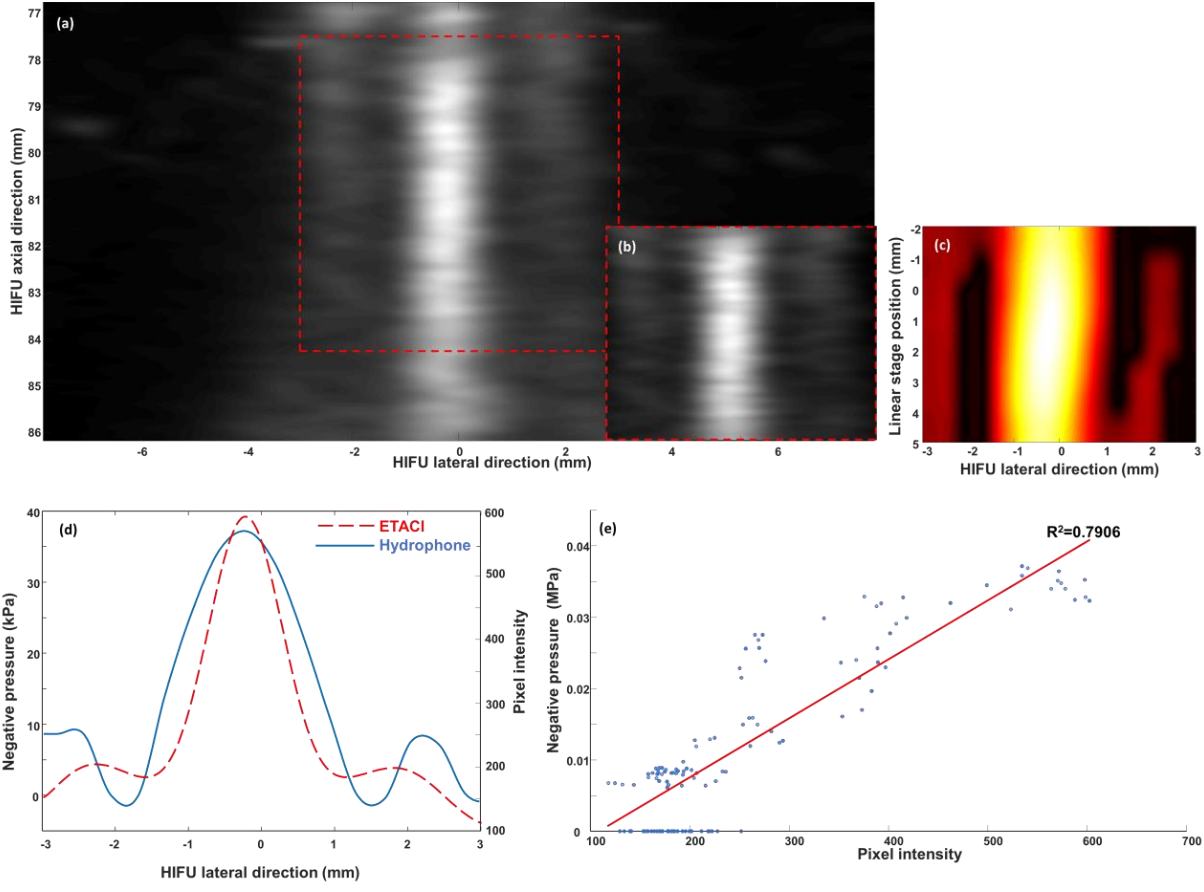

*Figure 6: High intensity focused ultrasound beam profile mapping in free field at 0.04 MPa peak negative pressure and hydrophone mapping of the same region. The red dotted line circumscribes the corresponding zoomed in area position on the overall field of view. (a) Overall field of view of the H-102 high intensity focused ultrasound pressure field measured with the ETACI modality. (b) Zoomed in area corresponding to the same region as the hydrophone measurement. (c) 2D central plane of the hydrophone pressure map. (d) Hydrophone mean negative pressure along the axial dimension and ETACI mean pixel intensity value along the axial dimension. The FWHM of ETACI and the hydrophone mapping are 1.50 mm and 1.88 mm respectively (e) Correlation between the pixel intensity of the ETACI technique and the negative pressure measurements of the hydrophone mapping.*

## Static Phantom characterization

To assess the capability to isolate the microbubble cavitation signal from a soft-tissue-like medium, a tissue-mimicking phantom was placed on the left side of the field of view. Figure 7(a-b) attests to our method's capability to allow tissue suppression and enhance the microbubble signal compared to a regular B-mode image of the same field of view. The phantom and the coupling cone artifact signals are mostly filtered out, resulting in only the pressure field mapping. The focal region of the pressure field seen in Figure 7(a) appears to have a focal dimension corresponding with the theoretical focal size (focal length 10.21 mm x focal width 1.37 mm at $F_0$). The intensity profile (Figure 7(c)) along the blue dotted line shows the lack of signal on the left side of the image where the phantom was standing, causing the absence of microbubble and thus a cavitation signal. This demonstrates that without microbubbles acting as local pressure sensors, the pressure field cannot be characterized in those regions.

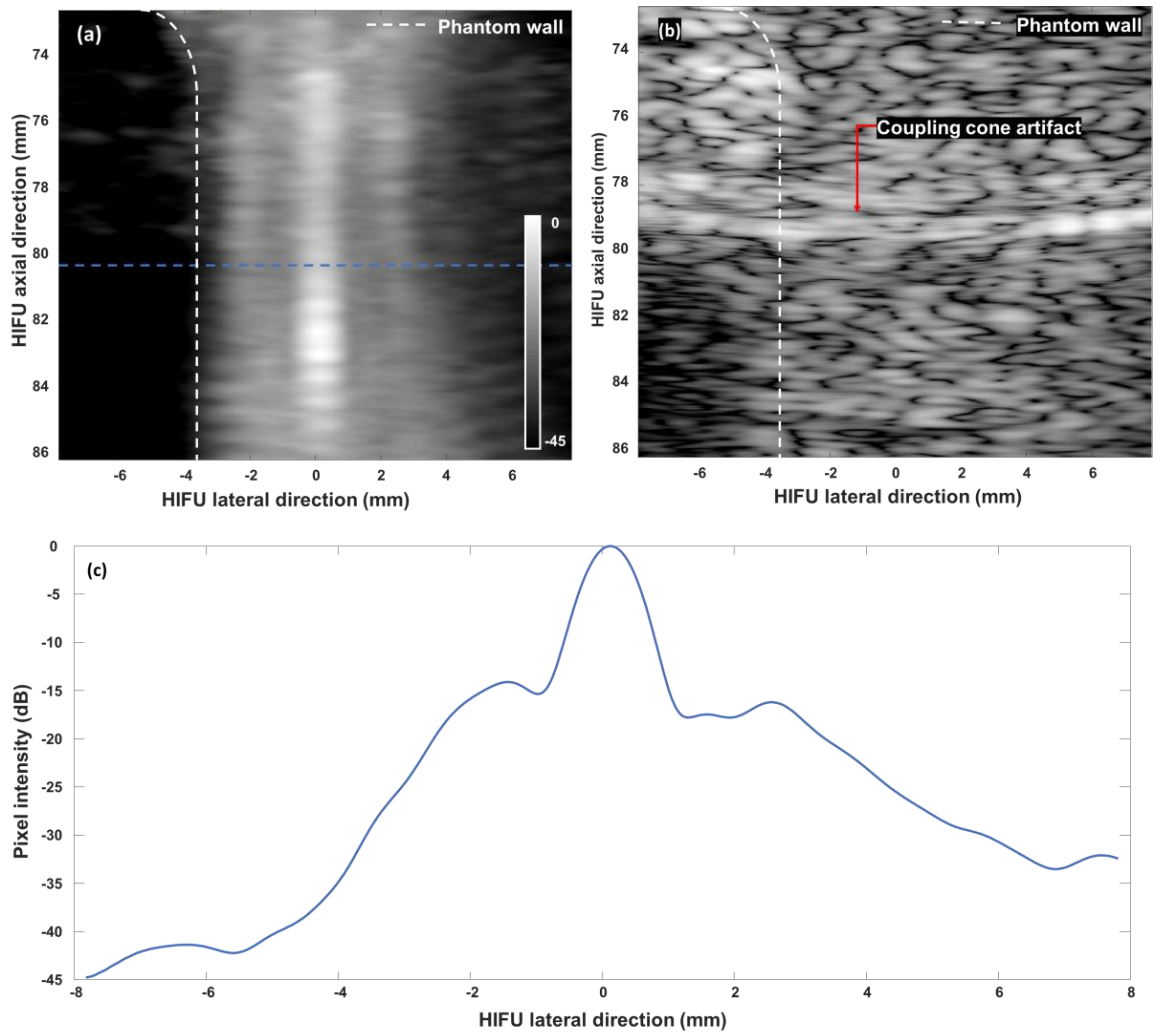

*Figure 7: HIFU pressure field and tissue suppression characterization. The white dotted line circumscribes the outline of Blue Phantom wall. (a) ETACI of the medium containing the free-floating microbubbles and the phantom. (b) B-mode of the same medium. (c) Pixel intensity profile in decibel along the blue dotted line crossing the focal field.*

## Flow phantom characterization

We evaluated the effect of gravity-generated flow on the ETACI modality. In order to have a better view of the pressure field inside the vessel, the imaging was performed along a longitudinal cut of a ~5 mm diameter vessel. As shown on Figure 8(a), the compounded ETACI pressure map characterizes the pressure field inside the vessel as microbubbles flow through it. Most of the signal is limited to the inside of the vessel and has the expected focal profile similar to the one in the free field (Figure 8(c)). The presence of speckle-like signals outside of the vessel can be explained by the presence of small gas pockets in the phantom medium.

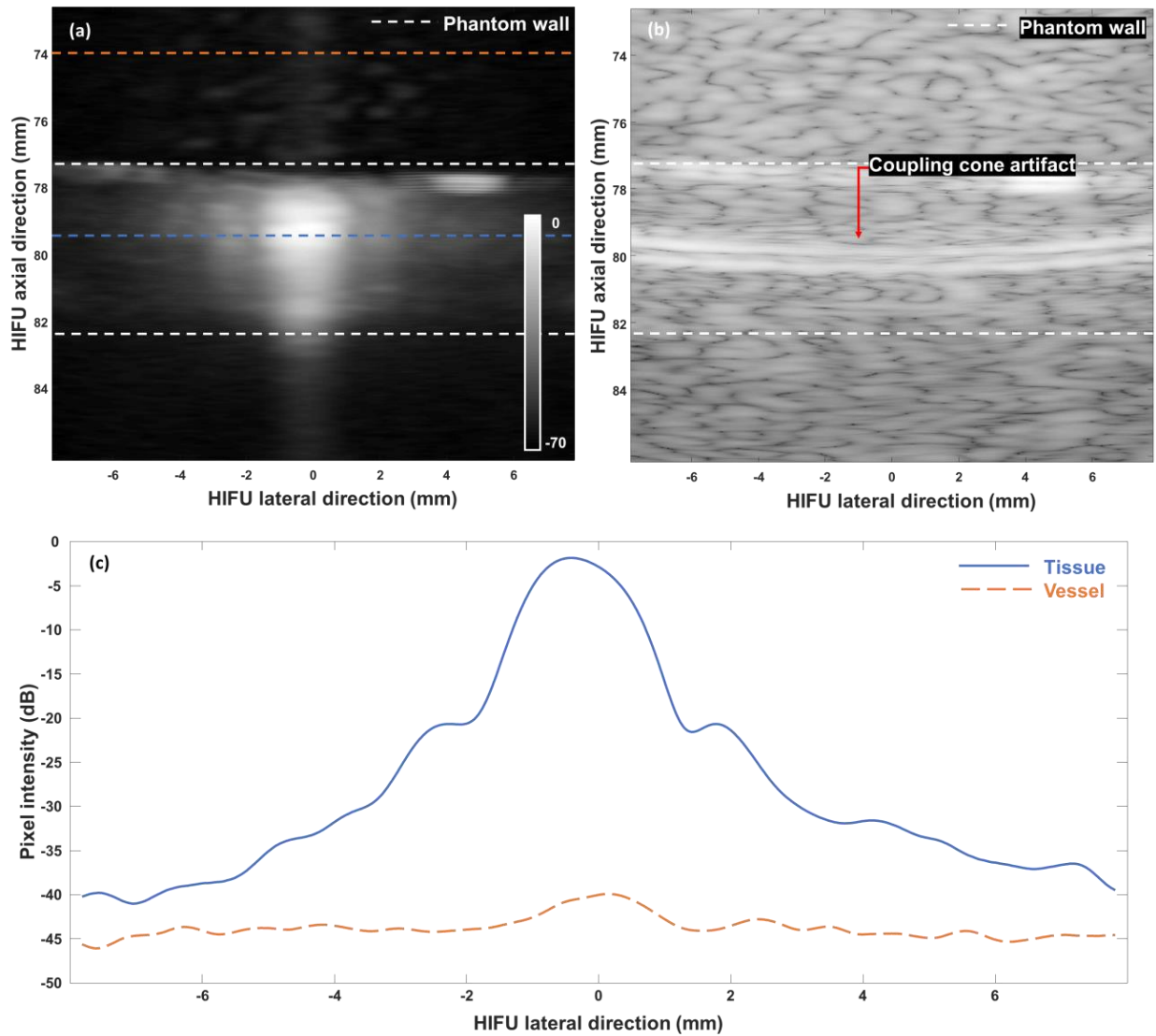

*Figure 8: Pressure field mapping over the longitudinal section of a vessel in a tissue mimicking flow phantom. (a) ETACI of the longitudinal section with gravity generated flow. (b) B-mode of the same medium. (c) Pixel intensity profile in decibels along the blue and orange dotted line situated on horizontal axis of the vessel and in the surrounding tissue respectively.*

## *In vivo* experiments

Figure 9 shows the initial proof of concept of the ETACI modality *in vivo* in the brain of an anesthetized mouse without craniotomy. Higher intensity regions in ETACI maps were observed to be overlapping with the corresponding BBB opened region as shown by autofluorescence of Evan's blue that had been injected in the tail vein simultaneously to microbubbles. However, ETACI signals were not detected for deeper structures in the burst treatment. The normalized integrated acoustic energy is shown in Figure 9(b-e) as yellow profiles indicate that the theoretical focal area might not be where the amount of acoustic energy is the highest, demonstrating the importance of this kind of imaging modality. This variation may be caused by beam deformation due to the skull, standing waves, and potentially, other biological effects. Further analysis of the average frequency spectrum profiles in the region of interest across all ensembles of a dataset shown in Figure 9(g) reveals an increase of base signal level between the burst treatment and the continuous one, as well as the presence of different frequency spikes compared to *in vitro* testing done at lower PNP. Ultra and sub-harmonic signals were present in the continuous treatment profile as well as an increase in the base signal. Accordingly, BBB opening has been confirmed on both sides by imaging of Evan's blue fluorescence (excitation at 620 nm, emission at 680 nm) while evidence for brain damage is provided by the presence of erythrocyte extravasation at the right cerebral hemisphere, which received a 0.40 MPa PNP continuous FUS treatment.

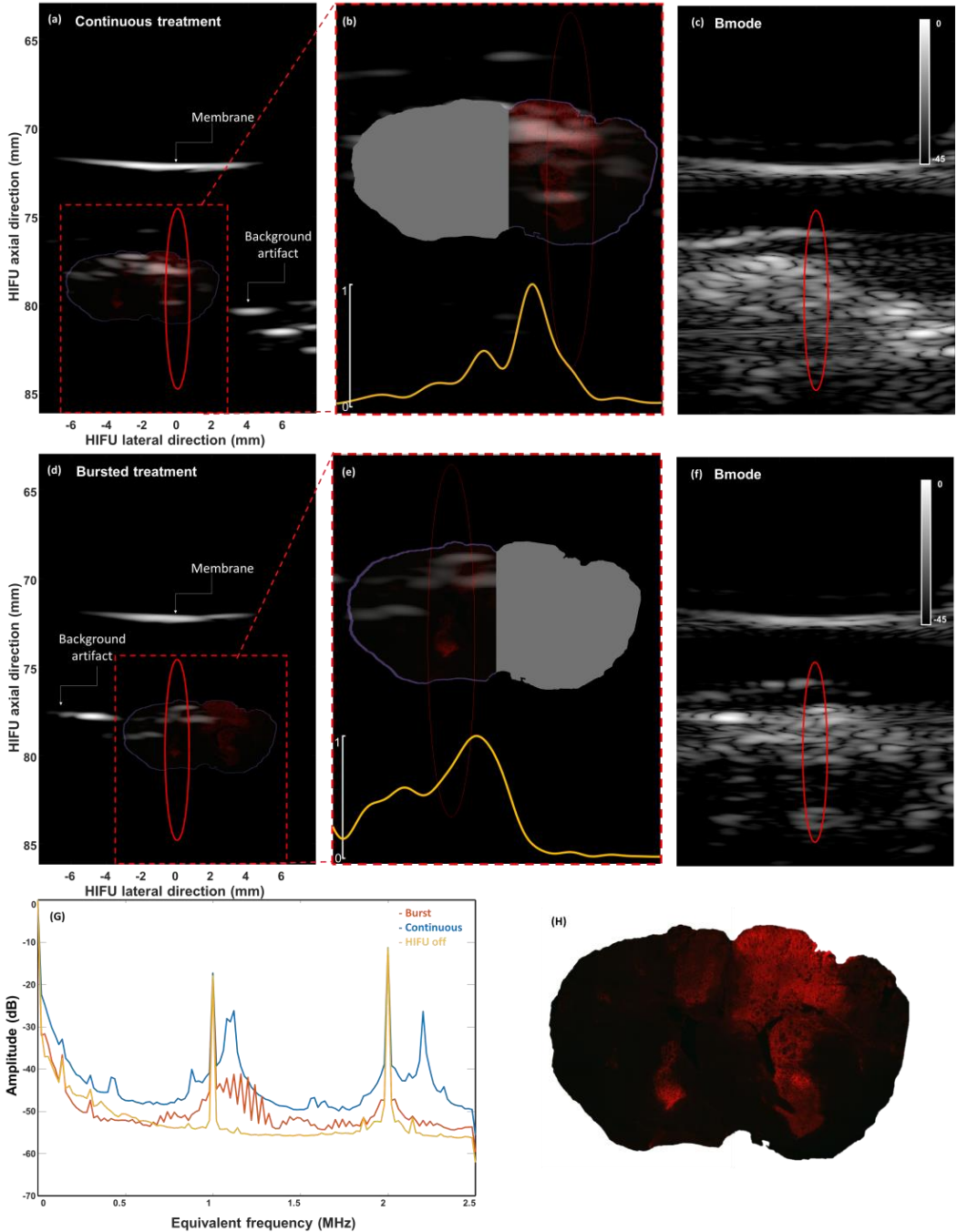

*Figure 9: Compounded pressure field mapping images of continuous and burst high intensity focused ultrasound blood-brain barrier opening treatment on the right and left side of the brain respectively. The dotted red rectangle shows the zoomed-in region and the red ellipse shows the theoretical focal size and position of the H-102 transducer in relation to the imaging probe. (a) Full-field of view of the compounded pressure field mapping of continuous high intensity focused ultrasound exposure treatment. (b) Overlay of the corresponding histology section with the zoomed-in compounded pressure field mapping. The yellow line shows the normalized acoustic energy integration along the axial dimension. (c) B-mode reference of the same field of view. (d) Full-field of view of the compounded pressure field mapping of continuous high intensity focused ultrasound exposure treatment. (e) Overlay of the corresponding histology section with the zoomed-in region in the compounded pressure field. The yellow line shows the normalized acoustic energy integration along the axial dimension. (f) B-mode reference of the same field of view. (g) Mean frequency spectrum profiles of the measured signal for burst and continuous treatment for mouse blood-brain barrier opening. (h) Fluorescence microscopy of a representative coronal brain section stained with Evans Blue dye.*

# Discussion

In this study, we have introduced a novel imaging sequence termed ETACI for the active mapping of cavitation events. By delaying plane wave emissions with a predetermined fraction of the HIFU period, we have shown the feasibility of mapping the pressure field (Figure 6), both using a static tissue phantom (Figure 7) and a flow phantom (Figure 8). We also show promising *in vivo* results, demonstrating the initial feasibility of ETACI in the context of BBB opening monitoring. *In vivo,* ETACI maps suggest that the method could be used to detect the position of the HIFU beam (Figure 9) and also signal the possible inertial cavitation events correlated with damage (Figure 9).

The frequency spectra obtained from ETACI display two prominent peaks, one at the driving frequency and the other at 1.12 MHz. The first peak was found to be associated with interpolation errors in the beamforming algorithm and could be further reduced by increasing the sampling frequency. The fact that the microbubble cavitation signal was shifted to 1.12 MHz instead of the driving frequency was unexpected. We advance the hypothesis that this could be caused by the interaction of the imaging pulse on the already oscillating microbubble, which could affect its measured oscillation period [47]. A similar phenomenon has also been raised by Muleki et al [52], which they called a *split spike*.

The pressure maps obtained in the free field showed a good correlation between ETACI's pixel intensity and the hydrophone's pressure measurements. These results indicate that ETACI could potentially be used as a rapid approach to transducer characterization since a single acquisition of a few seconds after mixing microbubbles in a water tank can be used to map entire fields of view. ETACI could also readily be generalized to 3D imaging by using 2-D matrix arrays. Additionally, pressure maps could be generated in presence of a static tissue phantom and in presence of flow, indicating that the approach may also be applied in contexts where it is not possible to perform hydrophone measurements. Finally, the tissue filtering capabilities of ETACI could also be used in the context of contrast-enhanced ultrasound imaging.

Active pressure field mapping has been shown possible *in vivo* during continuous and burst HIFU-induced BBB opening. While approximate relationships between the focal and BBB opening regions were observed, the spatial resolution that was obtained was largely degraded when compared against phantom experiments. Frequency spectra (Figure 9(a)) also appear to show that we can detect inertial cavitation during continuous treatment via a higher broad-band signal compared to burst treatment where no noticeable erythrocyte extravasation was observed. The discrepancy of the pressure values between in vivo and in vitro measurements is a limitation to our study, as they should have been at the same order of magnitude. This will be thoroughly addressed in further studies.

It is important to note that the setup used in this study presented for proof of principle, may be optimized. For instance, the imaging probe is more than 75 mm away from the region of interest, and 88 elements are used instead of the full 128 element array, thus greatly reducing the resolution, especially in the elevational (azimuthal) direction. This limitation can have a dramatic effect on imaging quality, especially in vivo. The next steps toward optimization would include an imaging transducer that would be lowered into a larger diameter FUS transducer allowing the imaging probe to be closer to the region of interest and increasing the resolution. Moreover, as the two focal planes are coincident, *in vitro* pressure field assessment is possible. However, as we progress to in *vivo* applications, and biological effects must be taken into account, the focal area can shift out of the imaging plane. In order to resolve this issue, 3D imaging based on a 2-D matrix-array probe could be used with the same algorithm. An offset in the coplanarity between the two transducers at the setup point can result in a significant rotational offset between the two planes at the focal level. The expansion ratio of the Definity microbubbles used in this study is defined by multiple parameters, such as the frequency, shell type, size, and peak negative pressure. It would be interesting to see how the arrangement of these parameters could help improve the method. Finally, the sequence used *in vivo* for burst treatment was not synchronized to the bursts themselves. Triggering the imaging on FUS burst train output could improve the slow time signal reconstruction and optimize the data acquisition.

ETACI could bring several benefits in comparison to other methods. For example, as this method relies on the fundamental resonance frequency, the possible signal gain is better than using a nonlinear response, as more energy is included in the fundamental frequency band compared to nonlinear frequency bands. The nonlinear response also increases the risk of microbubble rupture. Furthermore, ETACI doesn't require frequency locking and could also be used with different transducer assortments. This novel modality, therefore, allows for different imaging probe frequencies and combinations, which can vary depending on the experimental or clinical application. Multiple limitations have yet to be addressed and, as these benefits have yet to be proven, a direct comparison with PAM will be made in future studies. ETACI is currently implemented offline, which is a current limitation of this method but could be implemented in real-time for future studies. Moreover, as previously mentioned, this method is only valid under low-pressure conditions, which adds up to the other limitations of this current study.

# Conclusion

We have shown the feasibility of monitoring a modulation pressure field with an active imaging setup with plane wave imaging through microbubbles cavitational events using ETACI. This method addresses some of the limitations of existing methods for monitoring responses during HIFU applications and could significantly improve the ability to visualize targets for effective blood-brain barrier opening.

# Ethical considerations

Appropriate animal handling was done in observation of the Canadian Council on Animal Care guidelines and in accordance with McGill University Animal Care Committee regulation (Animal use protocol #2001-4532).

# Competing interests

A provisional patent has been made in relation with the ultrasound imaging modality developed in this study.

We acknowledge the support of TransMedTech, IVADO, CIHR, NSERC and the CFI (38095 and 246916).

# References

[1] V. L. Feigin *et al.*, "Global, regional, and national burden of neurological disorders, 1990–2016: a systematic analysis for the Global Burden of Disease Study 2016," *The Lancet Neurology*, vol. 18, no. 5, pp. 459–480, May 2019, doi: 10.1016/S1474-4422(18)30499-X.

[2] N. J. Abbott and I. A. Romero, "Transporting therapeutics across the blood-brain barrier," *Mol Med Today*, vol. 2, no. 3, pp. 106–113, Mar. 1996, doi: 10.1016/1357-4310(96)88720-x.

[3] S. Ohtsuki, "New Aspects of the Blood–Brain Barrier Transporters; Its Physiological Roles in the Central Nervous System," *Biological & pharmaceutical bulletin*, vol. 27, pp. 1489–96, Nov. 2004, doi: 10.1248/bpb.27.1489.

[4] U. Kniesel and H. Wolburg, "Tight Junctions of the Blood–Brain Barrier," *Cell Mol Neurobiol*, vol. 20, no. 1, pp. 57–76, Feb. 2000, doi: 10.1023/A:1006995910836.

[5]  W. M. Pardridge, "Molecular biology of the blood-brain barrier," *Mol Biotechnol*, vol. 30, no. 1, pp. 57–69, May 2005, doi: 10.1385/MB:30:1:057.
[6]  W. M. Pardridge, "The blood-brain barrier: bottleneck in brain drug development," *NeuroRx*, vol. 2, no. 1, pp. 3–14, Jan. 2005, doi: 10.1602/neurorx.2.1.3.
[7]  R. Daneman and A. Prat, "The Blood–Brain Barrier," *Cold Spring Harb Perspect Biol*, vol. 7, no. 1, p. a020412, Jan. 2015, doi: 10.1101/cshperspect.a020412.
[8]  R. A. Kroll and E. A. Neuwelt, "Outwitting the blood-brain barrier for therapeutic purposes: osmotic opening and other means," *Neurosurgery*, vol. 42, no. 5, pp. 1083–1099; discussion 1099-1100, May 1998, doi: 10.1097/00006123-199805000-00082.
[9]  N. D. Doolittle *et al.*, "Safety and efficacy of a multicenter study using intraarterial chemotherapy in conjunction with osmotic opening of the blood-brain barrier for the treatment of patients with malignant brain tumors," *Cancer*, vol. 88, no. 3, pp. 637–647, Feb. 2000, doi: 10.1002/(sici)1097-0142(20000201)88:3.
[10] R. H. Bobo, D. W. Laske, A. Akbasak, P. F. Morrison, R. L. Dedrick, and E. H. Oldfield, "Convection-enhanced delivery of macromolecules in the brain," *Proc. Natl. Acad. Sci. U.S.A.*, vol. 91, no. 6, pp. 2076–2080, Mar. 1994, doi: 10.1073/pnas.91.6.2076.
[11] S. Joshi *et al.*, "Inconsistent blood brain barrier disruption by intraarterial mannitol in rabbits: implications for chemotherapy," *J Neurooncol*, vol. 104, no. 1, pp. 11–19, Aug. 2011, doi: 10.1007/s11060-010-0466-4.
[12] X. Nassif, S. Bourdoulous, E. Eugène, and P.-O. Couraud, "How do extracellular pathogens cross the blood–brain barrier?," *Trends in Microbiology*, vol. 10, no. 5, pp. 227–232, May 2002, doi: 10.1016/S0966-842X(02)02349-1.
[13] K. Hynynen, N. McDannold, N. Vykhodtseva, and F. A. Jolesz, "Noninvasive MR imaging-guided focal opening of the blood-brain barrier in rabbits," *Radiology*, vol. 220, no. 3, pp. 640–646, Sep. 2001, doi: 10.1148/radiol.2202001804.
[14] K. Hynynen, N. McDannold, N. A. Sheikov, F. A. Jolesz, and N. Vykhodtseva, "Local and reversible blood–brain barrier disruption by noninvasive focused ultrasound at frequencies suitable for trans-skull sonications," *NeuroImage*, vol. 24, no. 1, pp. 12–20, Jan. 2005, doi: 10.1016/j.neuroimage.2004.06.046.
[15] N. Sheikov, N. McDannold, S. Sharma, and K. Hynynen, "Effect of focused ultrasound applied with an ultrasound contrast agent on the tight junctional integrity of the brain microvascular endothelium," *Ultrasound Med Biol*, vol. 34, no. 7, pp. 1093–1104, Jul. 2008, doi: 10.1016/j.ultrasmedbio.2007.12.015.
[16] G. Samiotaki and E. E. Konofagou, "Dependence of the Reversibility of Focused-Ultrasound-Induced Blood–Brain Barrier Opening on Pressure and Pulse Length In Vivo," *IEEE Trans Ultrason Ferroelectr Freq Control*, vol. 60, no. 11, pp. 2257–2265, Nov. 2013, doi: 10.1109/TUFFC.2013.6644731.
[17] N. McDannold, N. Vykhodtseva, and K. Hynynen, "Targeted disruption of the blood–brain barrier with focused ultrasound: association with cavitation activity," *Phys. Med. Biol.*, vol. 51, no. 4, pp. 793–807, Jan. 2006, doi: 10.1088/0031-9155/51/4/003.
[18] Y.-S. Tung, F. Vlachos, J. A. Feshitan, M. A. Borden, and E. E. Konofagou, "The mechanism of interaction between focused ultrasound and microbubbles in blood-brain barrier opening in mice," *J Acoust Soc Am*, vol. 130, no. 5, pp. 3059–3067, Nov. 2011, doi: 10.1121/1.3646905.
[19] M. J. K. Blomley, J. C. Cooke, E. C. Unger, M. J. Monaghan, and D. O. Cosgrove, "Microbubble contrast agents: a new era in ultrasound," *BMJ*, vol. 322, no. 7296, pp. 1222–1225, May 2001.
[20] P. THO, R. MANASSEH, and A. Ooi, "Cavitation microstreaming patterns in single and multiple bubble systems," *Journal of Fluid Mechanics*, vol. 576, pp. 191–233, Apr. 2007, doi: 10.1017/S0022112006004393.

[21] P. Marmottant and S. Hilgenfeldt, "Controlled vesicle deformation and lysis by single oscillating bubbles," *Nature*, vol. 423, no. 6936, Art. no. 6936, May 2003, doi: 10.1038/nature01613.
[22] H. Chen, W. Kreider, A. A. Brayman, M. R. Bailey, and T. J. Matula, "Blood Vessel Deformations on Microsecond Time Scales by Ultrasonic Cavitation," *Phys. Rev. Lett.*, vol. 106, no. 3, p. 034301, Jan. 2011, doi: 10.1103/PhysRevLett.106.034301.
[23] N. Ioritani *et al.*, "Renal tissue damage induced by focused shock waves," *AIP Conference Proceedings*, vol. 208, no. 1, pp. 185–190, Jul. 1990, doi: 10.1063/1.39438.
[24] Y. Lin *et al.*, "Effect of acoustic parameters on the cavitation behavior of SonoVue microbubbles induced by pulsed ultrasound," *Ultrasonics Sonochemistry*, vol. 35, pp. 176–184, Mar. 2017, doi: 10.1016/j.ultsonch.2016.09.016.
[25] S. Zhang *et al.*, "Dynamic changes of integrated backscatter, attenuation coefficient and bubble activities during high-intensity focused ultrasound (HIFU) treatment," *Ultrasound Med Biol*, vol. 35, no. 11, pp. 1828–1844, Nov. 2009, doi: 10.1016/j.ultrasmedbio.2009.05.003.
[26] M. T. Burgess, I. Apostolakis, and E. E. Konofagou, "Power cavitation-guided blood-brain barrier opening with focused ultrasound and microbubbles," *Phys Med Biol*, vol. 63, no. 6, p. 065009, 15 2018, doi: 10.1088/1361-6560/aab05c.
[27] A. F. Prokop, A. Soltani, and R. A. Roy, "Cavitational mechanisms in ultrasound-accelerated fibrinolysis," *Ultrasound Med Biol*, vol. 33, no. 6, pp. 924–933, Jun. 2007, doi: 10.1016/j.ultrasmedbio.2006.11.022.
[28] V. I. Ilyichev, V. L. Koretz, and N. P. Melnikov, "Spectral characteristics of acoustic cavitation," *Ultrasonics*, vol. 27, no. 6, pp. 357–361, Nov. 1989, doi: 10.1016/0041-624X(89)90034-6.
[29] C. C. Coussios, C. H. Farny, G. Ter Haar, and R. A. Roy, "Role of acoustic cavitation in the delivery and monitoring of cancer treatment by high-intensity focused ultrasound (HIFU)," *International Journal of Hyperthermia*, vol. 23, no. 2, pp. 105–120, Jan. 2007, doi: 10.1080/02656730701194131.
[30] T. Ilovitsh *et al.*, "Enhanced microbubble contrast agent oscillation following 250 kHz insonation," *Scientific Reports*, vol. 8, no. 1, Art. no. 1, Nov. 2018, doi: 10.1038/s41598-018-34494-5.
[31] C. Simon, P. VanBaren, and E. S. Ebbini, "Two-dimensional temperature estimation using diagnostic ultrasound," *IEEE Transactions on Ultrasonics, Ferroelectrics, and Frequency Control*, vol. 45, no. 4, pp. 1088–1099, Jul. 1998, doi: 10.1109/58.710592.
[32] T. N. Nguyen, M. N. Do, and M. L. Oelze, "Visualization of the Intensity Field of a Focused Ultrasound Source In Situ," *IEEE Transactions on Medical Imaging*, vol. 38, pp. 124–133, 2019, doi: 10.1109/TMI.2018.2857481.
[33] G. Pinton, J.-F. Aubry, E. Bossy, M. Muller, M. Pernot, and M. Tanter, "Attenuation, scattering, and absorption of ultrasound in the skull bone," *Medical Physics*, vol. 39, no. 1, pp. 299–307, 2012, doi: 10.1118/1.3668316.
[34] F. J. Fry, "Transkull transmission of an intense focused ultrasonic beam," *Ultrasound in Medicine & Biology*, vol. 3, no. 2, pp. 179–184, Jan. 1977, doi: 10.1016/0301-5629(77)90069-2.
[35] M. A. O'Reilly, A. Muller, and K. Hynynen, "Ultrasound insertion loss of rat parietal bone appears to be proportional to animal mass at submegahertz frequencies," *Ultrasound Med Biol*, vol. 37, no. 11, pp. 1930–1937, Nov. 2011, doi: 10.1016/j.ultrasmedbio.2011.08.001.
[36] M. Gerstenmayer, B. Fellah, R. Magnin, E. Selingue, and B. Larrat, "Acoustic Transmission Factor through the Rat Skull as a Function of Body Mass, Frequency and Position," *Ultrasound Med Biol*, vol. 44, no. 11, pp. 2336–2344, 2018, doi: 10.1016/j.ultrasmedbio.2018.06.005.
[37] M. Gyongy and C.-C. Coussios, "Passive Spatial Mapping of Inertial Cavitation During HIFU Exposure," *IEEE Trans. Biomed. Eng.*, vol. 57, no. 1, pp. 48–56, Jan. 2010, doi: 10.1109/TBME.2009.2026907.

[38] M. Marx, P. Ghanouni, and K. Butts Pauly, "Specialized volumetric thermometry for improved guidance of MRgFUS in brain," *Magn Reson Med*, vol. 78, no. 2, pp. 508–517, 2017, doi: 10.1002/mrm.26385.

[39] K. R. Gorny, N. J. Hangiandreou, G. K. Hesley, B. S. Gostout, K. P. McGee, and J. P. Felmlee, "MR guided focused ultrasound: technical acceptance measures for a clinical system," *Phys. Med. Biol.*, vol. 51, no. 12, pp. 3155–3173, Jun. 2006, doi: 10.1088/0031-9155/51/12/011.

[40] Y.-S. Tung, F. Vlachos, J. J. Choi, T. Deffieux, K. Selert, and E. E. Konofagou, "In vivo transcranial cavitation threshold detection during ultrasound-induced blood-brain barrier opening in mice," *Phys Med Biol*, vol. 55, no. 20, pp. 6141–6155, Oct. 2010, doi: 10.1088/0031-9155/55/20/007.

[41] S.-Y. Wu *et al.*, "Efficient Blood-Brain Barrier Opening in Primates with Neuronavigation-Guided Ultrasound and Real-Time Acoustic Mapping," *Sci Rep*, vol. 8, no. 1, pp. 1–11, May 2018, doi: 10.1038/s41598-018-25904-9.

[42] Z. Xu, Z. Fan, T. L. Hall, F. Winterroth, J. B. Fowlkes, and C. A. Cain, "Size Measurement of Tissue Debris Particles Generated from Pulsed Ultrasound Cavitational Therapy – Histotripsy," *Ultrasound Med Biol*, vol. 35, no. 2, pp. 245–255, Feb. 2009, doi: 10.1016/j.ultrasmedbio.2008.09.002.

[43] J. H. Song, K. Johansen, and P. Prentice, "Covert cavitation: Spectral peak suppression in the acoustic emissions from spatially configured nucleations," *The Journal of the Acoustical Society of America*, vol. 141, no. 3, pp. EL216–EL221, Mar. 2017, doi: 10.1121/1.4977236.

[44] K. J. Haworth, K. B. Bader, K. T. Rich, C. K. Holland, and T. D. Mast, "Quantitative Frequency-Domain Passive Cavitation Imaging," *IEEE Trans Ultrason Ferroelectr Freq Control*, vol. 64, no. 1, pp. 177–191, Jan. 2017, doi: 10.1109/TUFFC.2016.2620492.

[45] A. Llera, T. Wolfers, P. Mulders, and C. F. Beckmann, "Inter-individual differences in human brain structure and morphology link to variation in demographics and behavior," *eLife*, vol. 8, p. e44443, Jul. 2019, doi: 10.7554/eLife.44443.

[46] M. Tanter and M. Fink, "Ultrafast imaging in biomedical ultrasound," *IEEE Transactions on Ultrasonics, Ferroelectrics, and Frequency Control*, vol. 61, no. 1, pp. 102–119, Jan. 2014, doi: 10.1109/TUFFC.2014.2882.

[47] E. Chérin *et al.*, "Radial Modulation Imaging of Microbubble Contrast Agents at High Frequency," *Ultrasound in Medicine & Biology*, vol. 34, no. 6, pp. 949–962, Jun. 2008, doi: 10.1016/j.ultrasmedbio.2007.11.017.

[48] R. Hansen and B. A. J. Angelsen, "SURF imaging for contrast agent detection," *IEEE Transactions on Ultrasonics, Ferroelectrics, and Frequency Control*, vol. 56, no. 2, pp. 280–290, Feb. 2009, doi: 10.1109/TUFFC.2009.1037.

[49] K. Radhakrishnan *et al.*, "Relationship between cavitation and loss of echogenicity from ultrasound contrast agents," *Phys. Med. Biol.*, vol. 58, no. 18, pp. 6541–6563, Sep. 2013, doi: 10.1088/0031-9155/58/18/6541.

[50] F. Li, D. Li, and F. Yan, "Improvement of Detection Sensitivity of Microbubbles as Sensors to Detect Ambient Pressure," *Sensors (Basel)*, vol. 18, no. 12, Nov. 2018, doi: 10.3390/s18124083.

[51] N. de Jong, A. Bouakaz, and P. Frinking, "Basic acoustic properties of microbubbles," *Echocardiography*, vol. 19, no. 3, pp. 229–240, Apr. 2002, doi: 10.1046/j.1540-8175.2002.00229.x.

[52] P. Muleki-Seya, K. Xu, M. Tanter, and O. Couture, "Ultrafast radial modulation imaging," *IEEE Trans. Ultrason., Ferroelect., Freq. Contr.*, pp. 1–1, 2019, doi: 10.1109/TUFFC.2019.2949046.

[53] J. Park *et al.*, "Synthesis of Laboratory Ultrasound Contrast Agents," *Molecules*, vol. 18, no. 10, pp. 13078–13095, Oct. 2013, doi: 10.3390/molecules181013078.